\documentclass[twocolumn,showpacs,preprintnumbers,amsmath,amssymb,superscriptaddress]{revtex4}

\usepackage{graphicx}
\usepackage{dcolumn}
\usepackage{amsmath}    
\usepackage{verbatim}   
\usepackage{color}      
\usepackage{subfigure}  
\usepackage{hyperref}   
\usepackage{bm}

\begin{document}


\title{Free induction decay of a superposition stored in a quantum dot
}

\author{A. J. Bennett}
\affiliation{Toshiba Research Europe Limited, Cambridge Research Laboratory,\\
208 Science Park, Milton Road, Cambridge, CB4 0GZ, U. K.}

\author{M. A. Pooley}
\affiliation{Toshiba Research Europe Limited, Cambridge Research Laboratory,\\
208 Science Park, Milton Road, Cambridge, CB4 0GZ, U. K.}
\affiliation{Cavendish Laboratory, Cambridge University,\\
J. J. Thomson Avenue, Cambridge, CB3 0HE, U. K.}

\author{R. M. Stevenson}
\affiliation{Toshiba Research Europe Limited, Cambridge Research Laboratory,\\
208 Science Park, Milton Road, Cambridge, CB4 0GZ, U. K.}

\author{I. Farrer}
\affiliation{Cavendish Laboratory, Cambridge University,\\
J. J. Thomson Avenue, Cambridge, CB3 0HE, U. K.}

\author{D. A. Ritchie}
\affiliation{Cavendish Laboratory, Cambridge University,\\
J. J. Thomson Avenue, Cambridge, CB3 0HE, U. K.}

\author{A. J. Shields}
\affiliation{Toshiba Research Europe Limited, Cambridge Research Laboratory,\\
208 Science Park, Milton Road, Cambridge, CB4 0GZ, U. K.}

\date{\today}%

\begin{abstract}
We study the free evolution of a superposition initialized with high
fidelity in the neutral-exciton state of a quantum dot. Read-out of
the state at later times is achieved by polarized photon detection,
averaged over a large number of cycles. By controlling the
fine-structure splitting (FSS) of the dot with a DC electric field
we show a reduction in the degree of polarization of the signal when
the splitting is minimized. In analogy with the ``free induction
decay'' observed in nuclear magnetic resonance, we attribute this to
hyperfine interactions with nuclei in the semiconductor. We
numerically model this effect and find good agreement with
experimental studies. Our findings have implications for storage of
superpositions in solid state systems, and for entangled photon pair
emission protocols that require a small value of FSS.
\end{abstract}


\maketitle 



Quantum effects are often masked by interactions with the
environment. A well-known example is found in magnetic resonance
spectroscopy. Typically, a radio-frequency pulse is used to prepare
the nuclear spin states, which then precess around the applied
magnetic field \cite{Bloch46}. The signal obtained from
simultaneously measuring the projection of all spins along some
direction perpendicular to the field displays oscillations which
appear to fall away with time in a process known as ``free induction
decay'' (FID). In part this FID is due to the intrinsic decoherence
of the spins, a so called $T_{2}$ process. But in the solid state
this is often masked by a faster decay in the signal which arises
from variations in the field, susceptibility and local environment
of the nuclei, which consequently precess at different rates.

Quantum science can now routinely probe single solid state quantum
systems, such as the spin of electrons trapped at color centers in
diamond \cite{Neumann10}(which has a weak phonon interaction) or
single spins in silicon \cite{Morello10}(in which the host lattice
has nuclear spin zero). On the other hand III-V semiconductors have
both phonon and nuclear interactions to contend with, but are
nonetheless interesting for their potential scalability and
miniaturization, particularly as sources of non-classical light
\cite{Michler00,StevensonNature06}.

One of the most studied systems are single InGaAs/GaAs quantum dots
as they have optically active states with well-understood selection
rules, allowing spin-photon conversion \cite{flissikowski,
Kroutvar04, Atature06, Yilmaz10, Boyer10} and optical control
\cite{Greilich06, Ladd10}. Interactions of these states with the
nuclear spins in the quantum dot has lead to a wealth of new
physics, such as ''dragging'' the energy of a transition as it
follows a resonant laser \cite{Xu09, Latta09} and nuclear-spin
switching \cite{Tartakovskii07}. Most of the literature on the
effect of nuclear spins in a quantum dot has been concerned with the
spin eigenstates of the charge-exciton transition
\cite{Maletinsky07, Xu09, Latta09, Fallahi10, Yilmaz10, Ladd10} or
in some cases the neutral exciton transition in applied external
magnetic field \cite{Gammon01, Tartakovskii07, Latta09}. Through
this work it has been shown that the hole has a hyperfine
interaction which is an order of magnitude weaker than that of the
electron \cite{Fallahi10,Chekovich11}. In contrast, our work is
focussed on the behavior of superpositions stored in the
neutral-exciton at zero external magnetic field. Study of this
system is motivated by its ability to act as a a photon-exciton
interface  which will find applications in storage and manipulation
of photons \cite{flissikowski, Kroutvar04, Boyer10} and also in the
its central role in the emission of entangled photon pairs from the
neutral-cascade \cite{StevensonNature06}. We show that the
fluctuating magnetic field of nuclei overlapping with the
wavefunction of the exciton have a pronounced effect on the time
variation of the stored superposition. In analogy with NMR where a
large array of spins are measured simultaneously, our experiment
probes a single quantum system repeatedly over many fluctuations,
averaging the signal to observe a similar FID. Controlling the
energy splitting of the exciton eigenstates we are able to study the
FID of stored superpositions in the dot. We find a faster FID is
observed at small values of splitting, a result of the randomly
varying nuclear field. Intriguingly, we also find that when the
splitting is increased the FID timescale tends to a finite value
that is similar for all dots.

\begin{figure}[h]
\includegraphics[width = 80mm]{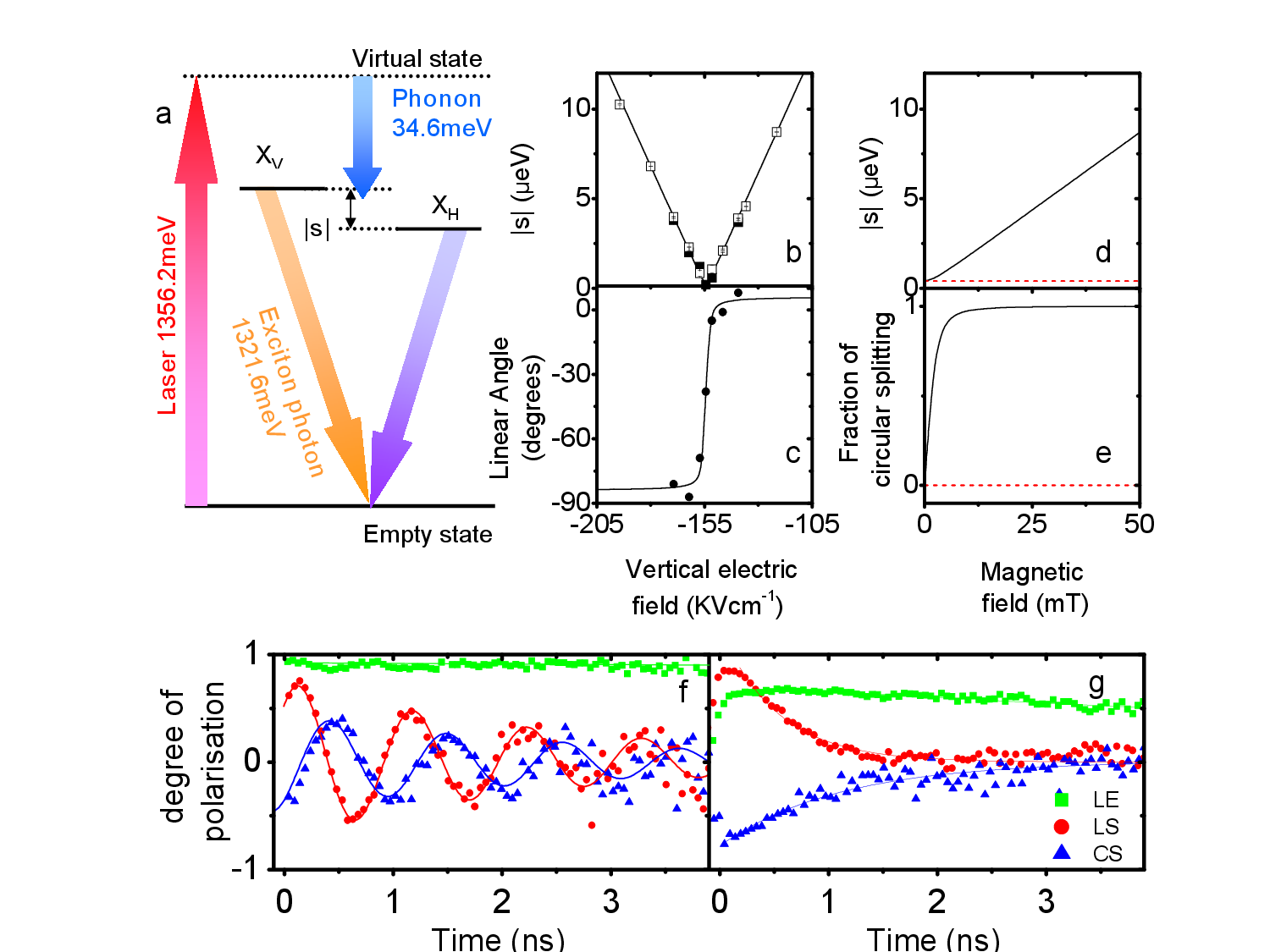} 
\caption{\label{Fig1} (a) Level diagram for the phonon-assisted
excitation of the neutral exciton states. Fine structure splitting
(b) and eigenstate orientation (relative to the laboratory frame)
measured in the spectral domain (filled data points) and temporal
domain (open data points) for a dot with a minimum splitting $s_{0}$
~ 0.4 $\mu eV$ as a function of vertical electric field. A fit
\cite{Bennett10} is included as a solid line. (d) Illustrates the
magnitude of the total FSS at $s_{0}$ as a function of vertical
(black) and in-plane (dashed) magnetic field used in the model. (e)
shows the resulting fraction of the total FSS due to splitting in
the circular basis as a function of magnetic field. (f)
Time-resolved degree of polarization for excitation and detection
along the linear eigenstate direction (LE, green), for the linear
superposition direction (LS, red) and circular superposition
direction (CS, blue) when the dot has a splitting of 3.7 $\mu eV$ at
- 132 kV/cm. (g) the same measurement when the splitting is minimal
at -153 kV/cm.}
\end{figure}

We begin by discussing our experiments. A single InGaAs/GaAs quantum
dot is excited one LO-photon energy (32 meV) above the exciton
transition using a mode-locked laser running at 80 MHz. With this
excitation scheme there is no phase-coherence between the laser and
the initialized state and the laser can be spectrally filtered from
the signal. However a coherent superposition between the populations
of the eigenstates is created, mapping the polarization of the laser
directly onto the Bloch-sphere of the solid-state exciton spin
\cite{flissikowski, Boyer10}. When the exciton later radiatively
decays the emitted light is passed to a fast silicon avalanche
photodiode (response time below 100 ps) and polarized time-resolved
data is acquired using counting electronics synchronized to the
driving laser. Traces are acquired in tens of seconds with
measurements parallel and orthogonal to the polarization of the
laser. This is carried out in all three measurement bases defined by
the linear eigenstates (LE), the linear superposition (LS) and
circular superposition (CS). Each pair is used to calculate
the``degree of polarization'' (the difference in the traces divided
by the sum) for that basis, such as the data shown in Figure
\ref{Fig1}f. The degree of polarization displays a noise which
increases with time, $t$, as the signal falls with a radiative
lifetime of 1.3ns. However, a good fit to the first 4ns of data can
be obtained using a least squares-fitting algorithm, to a function
of the form $\propto \sin (|s|(t-t_{0})/\hbar). \exp(-t/\tau_{FID})
$, where the term in $\tau_{FID}$ approximates an exponential decay
to the FID signal and $t_{0}$ describes the phase of the
oscillations. $s$ is the fitted fine-structure splitting (FSS)
between the two eigenstates. In the absence of a magnetic field,
anisotropy in the shape and strain of the dot, plus a contribution
from crystal asymmetry leads to a finite splitting between linear
eigenstates, $|s_{linear}|$, aligned with the crystal axes
 \cite{StevensonNature06}.

Our samples contain dots located in the center of a $p-i-n$ diode
with $Al_{0.75}Ga_{0.25}As$ barriers on either side, which allow
vertical electric fields to change the FSS \cite{Bennett10}. We
observe a minimum value of FSS, $|s_{0}|$, which varies from dot to
dot. Figure \ref{Fig1} shows experimental data from the exciton
state of one quantum dot with $|s_{0}| < 0.4 \mu eV$ at -153 kV/cm.
Away from the minimum value of FSS the LE are horizontally and
vertically linearly polarized in the laboratory frame, as is well
known for dots of this type. This is evidenced in Figure \ref{Fig1}f
which shows that excitation with light aligned with this LE retains
a high degree of polarization over the measurement. When the exciton
is excited in a superposition the finite value of FSS leads to
quantum beats in the intensity measured in a polarized measurement
with a period of $|s|/\hbar$ \cite{Haroche73, flissikowski,
Boyer10}. In the example shown in Figure \ref{Fig1}f $|s| = 3.7 \mu
eV$, and the degree of polarization decays with a lifetime of 3.0
ns. The envelope of the degree of polarization for the CS and LS
bases appear to degrade at a greater rate than the signal for the
linear eigenstate.

When the FSS is reduced to the minimum value, $|s_{0}|$, the
eigenstates of the exciton are rotated to be diagonal and
anti-diagonal in the laboratory frame \cite{Bennett10} (Figure
\ref{Fig1}c). Again, the degree of polarization from the LE
measurement decays at a rather slow rate. However, in the LS and CS
bases we expect to see an absence of oscillations and a similarly
slow decay rate, but instead Figure \ref{Fig1}g shows the degrees of
polarization both fall to zero with decay time of 0.5 ns. It appears
the state has decohered, but as we shall show this is in fact a
consequence of the hyperfine interactions with the nuclei in the
sample.

The effect of a magnetic field on the eigenstates of the exciton has
been well studied in other publications \cite{Bayer99, Stevenson06,
Bayer1999B} and the known behavior is plotted in Figure
\ref{Fig1}d-e. For an externally applied vertical magnetic field
(Faraday geometry) there is a splitting introduced in the circular
basis of $s_{circ} = g_{X}\mu_{B}B$, where $g_{X}$ is the exciton
g-factor (measured to be 3.0), $\mu_{B}$ the Bohr magneton and $B$
the applied field. This circular-basis splitting must be added in
quadrature to find the total splitting, $s = \sqrt(s_{circ}^{2} +
s_{linear}^{2})$ . The eigenstates now become predominantly
circular, with the fraction $s_{circ}/s$ tending to unity as shown
in \ref{Fig1}e. In contrast when the magnetic field is applied in
the plane of the sample (Voigt geometry) the splitting increases
only marginally according to $s = s_{linear} + \kappa B^{2}$ where
$\kappa$ is of the order of 1-2 $\mu eV T^{-2}$
\cite{StevensonNature06, Stevenson06, Bayer1999B}. For fields of up
to 50mT this results in change in fine-structure splitting too small
to measure.

\begin{figure}[h]
\includegraphics[width=80mm]{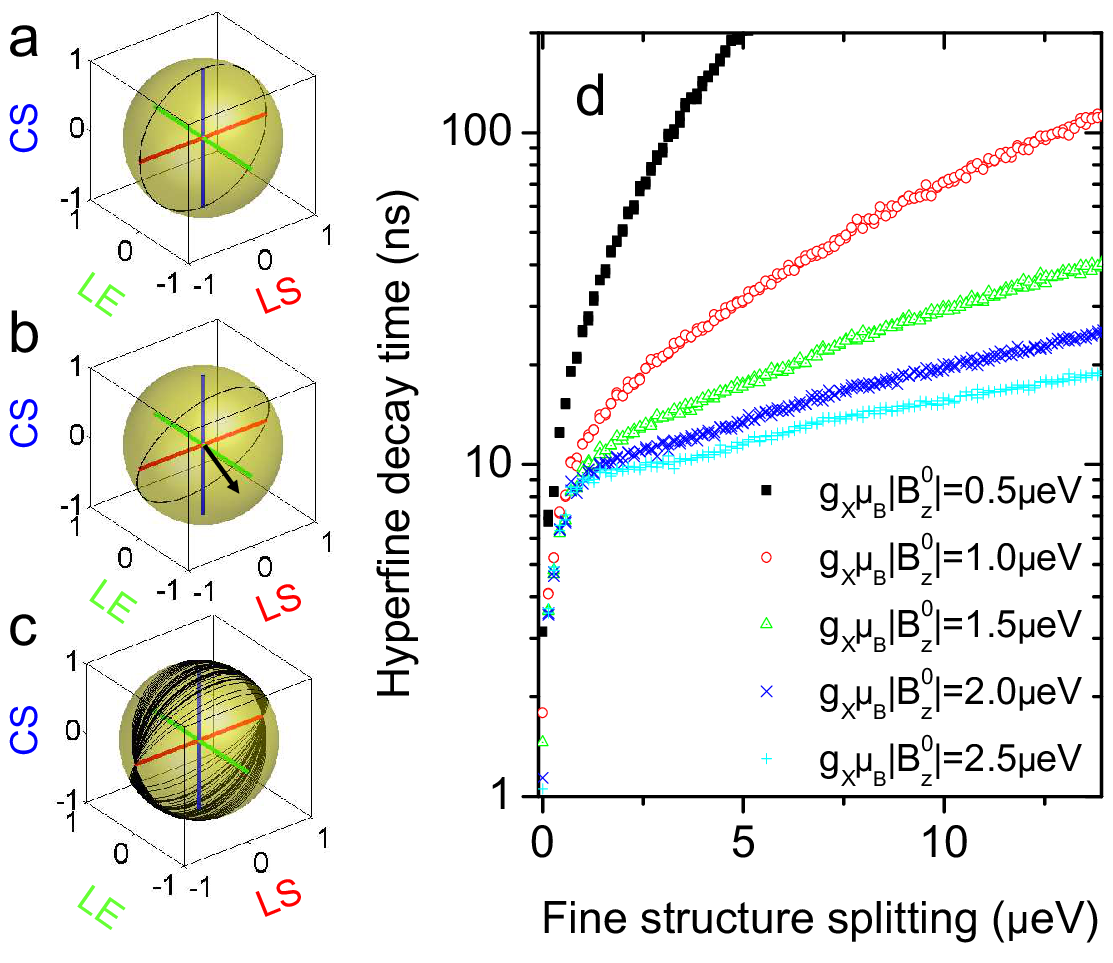}
\caption{\label{Fig2} (a) Bloch sphere for a qubit stored in the
quantum dot with linear-basis splitting only. The state precesses
around the Linear Eigenstate (LE) axis shown in green. (b) when a
finite magnetic field is present the eigenstate is slightly shifted
and the splitting is increased to $|S_{Ni}|$. The superposition now
evolves around the axis at a slightly increased rate
$|S_{Ni}|/\hbar$, and at a different angle. (c) when the
distribution of fields and angles that can occur is considered it
can be seen there are many possible paths. Performing Monte-Carlo
simulations for 20,000 values of field, we can see the
Free-induction decay of the degree of polarization due to this
effect varies as shown in (d). Calculations were performed for
different values of the initial linear splitting $|s_{linear}|$, and
gaussian-widths to the magnitude of field fluctuations, in units of
$g_{X} \mu_{B} B_{Ni}$.}
\end{figure}

Figure \ref{Fig2}a shows the path of a superposition on the Bloch
sphere for a pair of states with linear-basis splitting. The state
precesses around the linear eigenstate (LE) axis and polarized
measurements determine the averaged projection of the state along
that particular axis as a function of time. We then assume that
there is variable nuclear field which has some normal distribution
in its magnitude and random orientation. We consider only the
projection of the field $B_{Ni}$ at some time along the vertical
direction, leading to FSS in the circular basis $s_{circ} = g_{X}
\mu_{B} B_{Ni} \cos(\phi_{Ni})$, where $\phi_{Ni}$ the angle between
the field and the growth direction \cite{Bayer99}. We approximate
the nuclear spin interaction as an effective magnetic field and
effective g-factor. However, to maintain generality we do not
specifically assign the electron and hole contributions to the
effective g-factor. The magnitude of the FSS is now $|S_{Ni}| =
\sqrt(s_{linear}^2 + (g_{X} \mu_{B} B_{Ni} \cos (\phi_{Ni}))^{2})$
\cite{Eble06}. The eigenstates of the exciton are rotated by an
angle $\theta = tan^{-1}(\frac{g_{X} \mu_{B} B_{Ni}
\cos(\phi_{Ni})}{s_{linear}})$. The state initialized then rotates
around this new eigenstate at a rate of $S_{Ni}/\hbar$, as shown in
Figure \ref{Fig2}b. To calculate the result of a measurement along
the LS, LE and CS directions we extract the projection of this path
along that direction.

\begin{subequations}
 \begin{equation}
 V_{LS} \propto  \cos (|S_{Ni}|t/\hbar).e^{-t/\tau_{HV}}
 \end{equation}
 \begin{equation}
 V_{LE} \propto  \sin (|S_{Ni}|t/\hbar). \sin (\phi_{Ni}).e^{-t/\tau_{HV}}
 \end{equation}
 \begin{equation}
 V_{CS} \propto -  \sin (|S_{Ni}|t/\hbar). \cos (\phi_{Ni}).e^{-t/\tau_{HV}}
 \end{equation}
\end{subequations}

In these equations we have also included an extra exponential
decoherence term, with a timescale $\tau_{HV}$, the origin of which
shall be discussed later.

In practice fluctuations in the nuclear field occur over a time
scale of milliseconds \cite{Bayer99, Maletinsky07} which ensures
that over a measurement many trajectories over the Bloch sphere are
sampled. To calculate the effect this has we carry out Monte-Carlo
simulations for 20,000 values of field magnitude and orientation
denoted $i$. The results in Figure \ref{Fig2}(d) show the timescale
of the FID due only to the hyperfine interaction (excluding the
effect of $\tau_{HV}$) for different values of the initial linear
splitting $|s_{linear}|$, as a function of the magnitude of the
nuclear field fluctuations, where the different curves refer to
varying Gaussian width of the distribution, in units of $g_{X}
\mu_{B} B_{Ni}$.

From this model we can infer a few interesting facts. $(1)$ The
nuclear field acts only to increase the total FSS, $|S_{Ni}|$. Thus
a temporal measurement of the FSS averaged over many $i$ should
always lead to a marginally greater value than determined from a
spectral measurement in the linear bases. Note that an averaging
over $i$ leads to a circular splitting much less than $\langle g_{X}
\mu_{B} |B_{Ni}|\rangle$ as the field orientation is random. In turn
this leads to an even smaller increase in the total splitting
because the linear and circular splitting are added in quadrature.
In practice, the error on the spectral measurement of FSS is greater
than 0.3 $\mu eV$ \cite{Bennett10} so this difference between the
temporal $|s|$ and spectral measurement of $|s_{linear}|$ cannot be
observed (Figure \ref{Fig1}b). $(2)$ From Figure \ref{Fig2}c it can
be seen that measurements of the free-induction decay will be
different for measurements along the LS and CS axes. Consider for
example the creation of a linear superposition, which is then
measured along the linear superposition axis: in this case the
measurement probes only the fluctuation in the magnitude of $S_{Ni}$
but not the orientation of the eigenstates. In contrast,
measurements projected along the CS axis are sensitive to both
$B_{Ni}$ and $\phi_{Ni}$ and thus decay at a different rate. $(3)$
Even when the excitation or measurement of the exciton is made in
the time-averaged eigenstate, some weak oscillations will be
observed as the state fluctuates into the circular basis. This is
observed in our data, for example in the measurement of the LE shown
in Figure \ref{Fig1}f. $(4)$ Finally, our model suggests that in the
case where read out of the quantum trajectory can be made faster
than the nuclear fluctuations this effect will be absent.

\begin{figure}[h]\includegraphics[width=50mm]{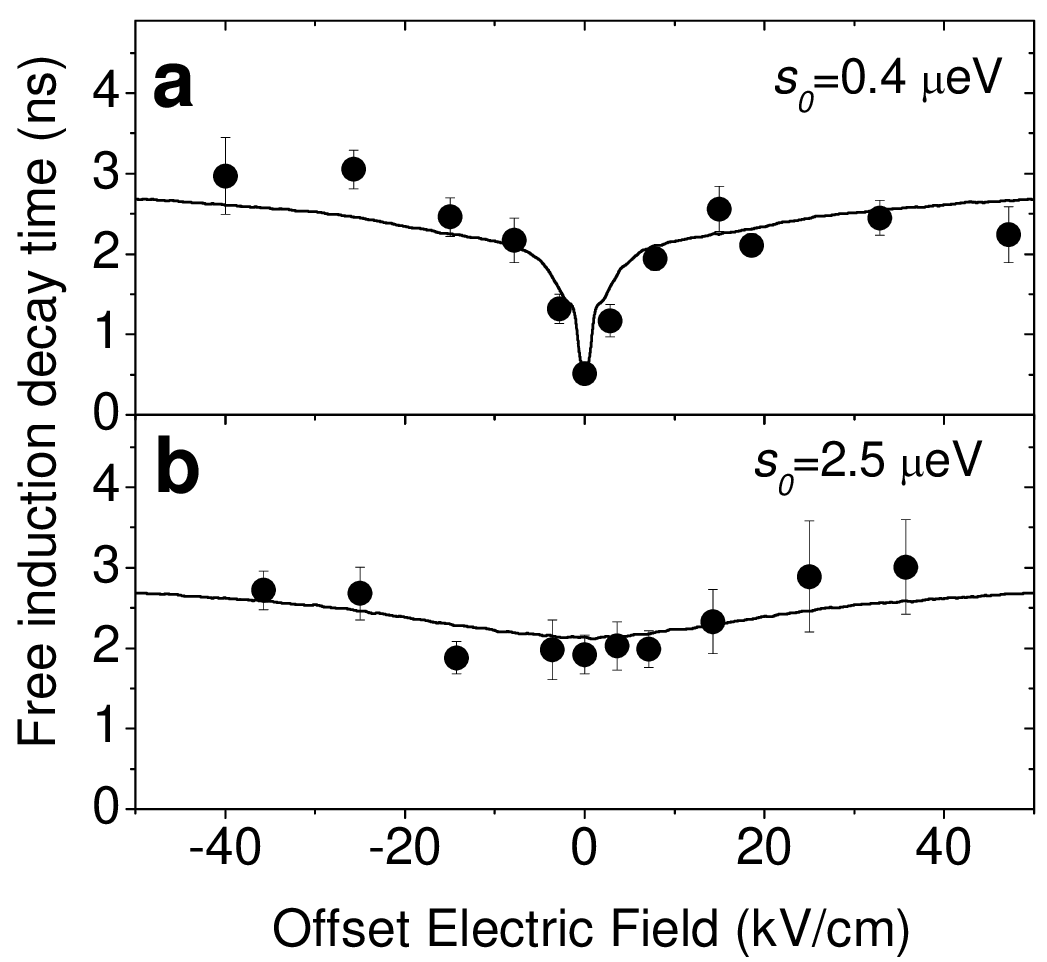}
\caption{\label{Fig3} Free induction decay time of a linear
superposition stored in the exciton state of a single quantum dot.
\textbf{a} For a dot with a minimum value of FSS of ~ 0.4 $\mu eV$.
\textbf{b} For a dot with a minimum value of FSS of 2.5 $\mu eV$}
\end{figure}

We have measured the rate of FID for a number of dots, as the FSS is
varied (Figure \ref{Fig3}). In all cases excitation with a low power
density was used to ensure spin-pumping of the nuclear field was not
occurring \cite{Tartakovskii07}. What we observe is that in the dot
with the lowest $|s_{0}|$ (Figure \ref{Fig1}) we see a rapid FID
when the FSS is small (Figure \ref{Fig3}a). The variation in the FID
time as a function of field (and FSS) is symmetric, suggesting is it
an effect dominated by the absolute value of the FSS, not by the
electric field. Measurements shown in Figure \ref{Fig3}b on a dot
where $|s_{0}| = 2.5 \mu eV$  show an accordingly greater FID time
at the minimum FSS.

In Figure \ref{Fig3} the solid line represents the FID time
extracted from our simulation for a Gaussian-width to the
distribution nuclear fields of $g_{X} \mu_{B} |B_{N}|$ = 2 $\mu eV$
and $\tau_{HV}$ = 3.0ns. Clearly, near $|s_{0}|$ the FID is
dominated by the nuclear fluctuations, suggesting the width of the
fluctuations in $|B_{N}|$ are of order 12 mT, and that several
10$^{3}$ nuclei are interacting with the exciton. At larger values
of splitting we see a decay determined by $\tau_{HV}$, which
describes the characteristic rate at which the $H$ and $V$
components of the superposition lose relative phase ( the ``cross
dephasing time'' \cite{stevenson2}, equivalent to the $T_{2} ^{*}$
in NMR.

We also observe that the fidelity of initializing a superposition is
maximized at $|s_{0}|$, but the fidelity of initializing an
eigenstate is minimized. This is because at $|s_{0}|$ fluctuations
in nuclear field have the greatest effect on the orientation of the
eigenstate on the Bloch-sphere. Conversely, exciting a superposition
using an optical transition with a finite line-width leads to a
jitter in the time at which the exciton is created. This acts to
reduce the initial degree of polarization when the coherent
oscillations are faster at large $|s|$. Thus the time-averaged
degree of polarization displays a maximum in the fidelity of
superposition as the FSS is tuned through $|s_{0}|$ \cite{kowalik}.
The width of this variation and the maximum degree of polarization
is determined by the fluctuations and dephasing processes that
control the free-induction decay, not the line-widths of the
individual exciton transitions.

The arguments presented here are also relevant to the case of
entangled photon pair generation, where decay of the biexciton
prepares the exciton in a given superposition \cite{stevenson2011}.
Many publications have stressed the importance of reducing the FSS
for a given dot to allow for the emission of entangled photon pairs
\cite{StevensonNature06}. This condition is not necessary when
temporal filtering can be applied to resolve the coherent
oscillations of the exciton \cite{stevenson2}. In addition to these
considerations we stress here that the only necessary and sufficient
condition for entangled photon emission is that the $T_{2}$ decay of
the $X$ state not be faster than the radiative lifetime or detector
response time, which ever is lower. If this is the case the fidelity
of entangled photon pair emission cannot be re-claimed through
temporal filtering.

In conclusion, we have shown the effects of nuclear field
fluctuations in the semiconductor environment cannot be ignored when
considering the evolution of a superposition trapped in a single
quantum dot. These fluctuations lead to a reduction in the degree of
polarization for the superposition, in a process closely analogous
to the free-induction decay of spins in nuclear magnetic resonance.
An interesting avenue for future research would be the combination
of our measurements and the use of spin-pumping to partially align
the nuclear field.

\section{Acknowledgements}

This work was partly supported by the EU through the Integrated
Project QESSENSE, ITN Spin-optronics and EPSRC.


\end{document}